\documentstyle[12pt,aasms4]{article}
\newcommand{\postscript}[2]
 {\setlength{\epsfxsize}{#2\hsize}
  \centerline{\epsfbox{#1}}}
\def\tempest%
{\begin{array}{ccc}
1 & 1 & 1 \\
1 & 1 & 1 \\
4 & 3 & 8
\end{array}}

\begin{document}

\title{A Comparison of the Intrinsic Shapes of \\
	Two Different Types of Dwarf Galaxies: \\
	Blue Compact Dwarfs and Dwarf Ellipticals}
\bigskip
\author{Eon-Chang Sung}
\smallskip
\affil{Korea Astronomy Observatory, Taejon, Korea 305-348 \\
       ecsung@hanul.issa.re.kr,}
\authoremail{ecsungi@hanul.re.kr}
\bigskip

\author{Cheongho Han}
\affil{Department of Astronomy \& Space Science, \\
       Chungbuk National University, Cheongju, Korea 361-763 \\
       cheongho@astro-3.chungbuk.ac.kr,}
\authoremail{cheongho@astro.chungbuk.ac.kr}
\bigskip

\author{Barbara S. Ryden}
\affil{Department of Astronomy, The Ohio State University\\
       174 West 18th Avenue, Columbus, OH 43210 \\
       ryden@astronomy.ohio-state.edu,}
\authoremail{ryden@astronomy.ohio-state.edu}

\bigskip
\author{Mun-Suk Chun}
\smallskip
\affil{Department of Astronomy, \\
       Yonsei University, Seoul, Korea 120-749 \\
       mschun@galaxy.yonsei.ac.kr,}
\authoremail{mschun@galaxy.yonsei.ac.kr}
\bigskip

\bigskip
\author{Ho-Il Kim}
\smallskip
\affil{Korea Astronomy Observatory, Taejon, Korea 305-348 \\
       hikim@hanul.issa.re.kr.}
\authoremail{hikim@hanul.re.kr}
\bigskip

\begin{abstract}

We measure the apparent shapes for a sample of 62 blue
compact dwarf galaxies (BCDs), and compare them with the
apparent shapes for a sample of 80 dwarf elliptical
galaxies (dEs). The BCDs are flatter, on average,
than the dEs, but the difference is only marginally significant.
We then use both non-parametric and parametric techniques
to determine possible distributions of intrinsic shapes for
the BCDs. 
The hypothesis that BCDs are oblate spheroids
can be ruled out with a high confidence level ($> 99\%$), but
the hypothesis that they are prolate spheroids cannot be excluded.
The apparent shapes of BCDs are totally consistent with
the hypothesis that they are triaxial ellipsoids.
If the intrinsic axis ratios, $\beta$ and $\gamma$, are distributed
according to a Gaussian with means $\beta_0$ and $\gamma_0$ and
standard deviation $\sigma$, we find the best-fitting 
distribution for BCDs has $(\beta_0,\gamma_0,\sigma)= (0.66,0.55,0.16)$,
while that for dEs has $(\beta_0,\gamma_0,\sigma)= (0.85,0.64,0.24)$.
Our results are consistent with the hypothesis that BCDs
have a close evolutionary relation with dEs.

\end{abstract}

\vskip100mm

\centerline{submitted to {\it The Astrophysical Journal}: September 18, 1997}
\centerline{Preprint: CNU-A\&SS-04/97}
\clearpage

\section{Introduction}
Understanding the formation and evolution
of stellar systems like galaxies begins from the comparison of 
various observables.
Dwarf galaxies are the most numerous type of galaxy in the universe.
Nearly $80\%$ of the known Local Group galaxies are dwarfs, and the 
number density of dwarf galaxies might be several dozen 
times that of giant spirals
and ellipticals (Staveley-Smith, Davies, \& Kinman 1992).\markcite{
staveley1992}
The most common type of dwarf galaxy is the dwarf
elliptical (hereafter dE).
They have very smooth surface brightness profiles, 
which are typically nearly exponential
(Sandage \& Binggeli 1984; 
Caldwell \& Bothun 1987).\markcite{sandage1984, caldwell1987}
The dEs, like giant ellipticals, are found predominantly in groups and
clusters of galaxies.
Most dEs are symmetric and apparently relaxed with very uniform colors 
indicating that they are dominated by old stars (James 1994).\markcite{
james1994}
The best-studied dwarf ellipticals are those in the Virgo cluster,
cataloged by Sandage \& Binggeli (1984) and
Binggeli, Sandage, \& Tammann (1985).\markcite{binggeli1985}

A smaller group of dwarf galaxies are classified as
``blue compact dwarfs'' (hereafter BCDs).
In contrast to dEs, 
BCDs are undergoing intense bursts of star formation
which give birth to a large number of massive stars,
as evidenced by their blue $UBV$ colors. 
These very young hot stars ionize the interstellar medium, 
producing high-excitation super-giant HII regions 
(Thuan 1991; Thuan, Izotov, \& Lipovetsky 1995).\markcite{thuan, thuan1995}
As a result, the optical spectra of BCDs show strong narrow emission lines.
Their ultraviolet spectra show a steeply
rising continuum toward the blue, which is characteristic of 
OB stars (du Puy 1970; Searle \& Sargent 1972).\markcite{dupuy1970, 
searle1972}

In addition to these spectroscopic and spectrophotometric observations, 
another important tool for the comparison of different
populations of dwarf galaxies is provided by morphological studies --
for instance, the study of intrinsic shapes based
on measured ellipticity distributions.
There have been several trials which have explored this point:
Caldwell (1983), Ichikawa, Wakamatsu, \& Okamura (1986),
Ichikawa (1989), Ferguson \& Sandage (1989),
and Ryden \& Terndrup (1994) for dEs and Staveley-Smith et al.\ (1992) 
for BCDs.\markcite{caldwell1983, ichikawa1986, ichikawa1989, 
ferguson1989, ryden1994, staveley1992}
However, due to different observational techniques, different
methods of measuring ellipticities, and different algorithms
for deriving intrinsic shapes,
direct comparison of the results might cause systematic errors.
In addition, some
previous studies suffered from additional systematic errors in 
eye-estimated flattening data (Fasano \& Vio 1991; 
Ferguson \& Binggeli 1994).\markcite{fasano1991, ferguson1994}
Among the above authors, only the analysis of Ryden \& Terndrup (1994) 
was based on CCD observations of a large data set.
The analysis of Staveley-Smith et al.\ (1992)\markcite{ryden1994, staveley}
was also based on CCD observations;
however, they determined the intrinsic shapes of
BCDs under the (probably erroneous) assumption that all BCDs
are oblate spheroids. In addition, the total size of their
sample is too small to draw any strong conclusion.
For studies of intrinsic shapes, a large homogeneous data
set is essential.

In this paper, we measure the ellipticities of 62 BCDs, and
compare the ellipticity distribution with that of a sample
of 80 dEs (Ryden \& Terndrup 1994; Ryden et al. 1997).\markcite{ryden1994,
ryden1997}
The ellipticity distribution of BCDs proves to be marginally different
from that of dEs.
We then determine the distribution of intrinsic shapes for
the BCDs, under various assumptions. 
First, we assume that BCDs are either all oblate or all prolate,
and use non-parametric analysis to find the best-fitting
distribution of intrinsic shapes.
Next, we assume that BCDs are triaxial, and use parametric
analysis to find permissible distributions of intrinsic shapes.
From these analyses, we find that the shapes of BCDs are well described 
either by triaxial ellipsoids or prolate spheroids;
the hypothesis that BCDs are all oblate is ruled out.
Similar results are obtained for dEs.
If the intrinsic axis ratios, $\beta$ and $\gamma$, are distributed
according to a Gaussian with means $\beta_0$ and $\gamma_0$ and
a common standard deviation of $\sigma$, we find the best-fitting 
intrinsic axis ratio
distribution of BCDs has $(\beta_0,\gamma_0,\sigma)= (0.66,0.55,0.16)$
with the KS probability of $P_{\rm KS} =0.99$,
while that of dEs has $(\beta_0,\gamma_0,\sigma)= (0.85,0.64,0.24)$
with $P_{\rm KS}=0.97$. 
Given the similarity of shape distributions, our results
are consistent with the hypothesis that BCDs have a close
evolutionary relation with dEs.

\section{Observations} 

To measure the apparent axis ratios,
we obtained images of 62 BCDs with the 1 m telescope at 
Mount Stromlo \& Siding Spring Observatory.
Eight observing sessions were carried out from 1993 April to 1994 March,
employing a $1024\times 1024$ Tektronics chip, which provides 
a field of view of $10'\times 10'$ with a pixel size 
of $0''\hskip-2pt .57 {\,\rm pixel}^{-1}$.
The exposure time varied from 300 s to 600 s depending on 
the brightnesses of individual galaxies.
We attempted to obtain the 4 (Bessell 1990)\markcite{bessell1990} 
$BVRI$ band images of individual BCDs.
However, due to limited observing time, we obtained full 4-band images
for only 35 out of 62 galaxies. For the other 
galaxies we obtained partial band images: 3 bands for
7 galaxies, 2 bands for another 7 galaxies, and only a single band image
for the remaining 13 galaxies.  
Therefore, the total number of images is 188.
The observed bands of each image are listed in Table 1.

We selected our sample of BCDs from various sources.  
We choose nearly half of our sample galaxies from 
the list of Thuan \& Martin (1981)\markcite{thuan1981};
and the rest are from 
Maza et al.\ (1991)\markcite{maza1991},
Bergvall \& Olofsson (1986)\markcite{bergvall1986},
Kunth \& Sargent (1986)\markcite{kunth1986},
Macalpine \& Williams (1981)\markcite{macalpine1981},
Fairall (1977)\markcite{fairall1977},
Terlevich et al.\ (1991)\markcite{terlevich1991},
Gondhaeker et al.\ (1984)\markcite{gondhaeker1984}, 
Wamsteker et al.\ (1985)\markcite{wamsteker1985}, and
Acker, Stenholm, \& Verson (1991)\markcite{acker1991}.
The galaxies in these sources are classified as BCDs according to 3 
major criteria:

\begin{minipage}{6.3in}
\begin{flushleft}
1.\ low luminosity, with $M_B \gtrsim -18$,

2.\ strong, sharp, narrow emission lines superimposed on a blue continuum, 

3.\ very compact, with diameters of a few kpc.
\end{flushleft}
\end{minipage}

\noindent
To meet the necessities of observation and data analysis,
we impose several additional criteria:

\begin{minipage}{6.3in}
\begin{flushleft}
4.\ relatively small recession velocities
($v \lesssim 4,000\ {\rm km\ s}^{-1}$ ),

5.\ relatively bright apparent magnitudes ($B < 16$),

6.\ relatively large angular sizes ($\gtrsim 20''$),

7.\ unperturbed outer isophotes.
\end{flushleft}
\end{minipage}

\noindent
We include conditions 4 and 5 to select galaxies 
bright enough to obtain high $S/N$ images with the available telescope.
Condition 6 is included because it is difficult
to measure ellipticities precisely for galaxies
of small angular size.
According to the classification scheme
of Loose \& Thuan (1986)\markcite{loose1986},
BDCs are classified into four types based on the regularity 
of the isophotal shapes of both the high surface brightness
inner region, where star formation has recently occurred,
and the low surface brightness outer region, which contains
only an older stellar population.
The first and most common type, comprising $\sim 70\%$ of BCDs,
is `iE', which combines irregular inner isophotes with accurately
elliptical outer isophotes. The `nE' type has elliptical
isophotes in both the inner and outer regions.
The `iI' type has irregular inner isophotes and outer
isophotes that deviate measureably from perfect ellipses.
Finally, `iO' galaxies do not show faint outer structures
at all; they appear to contain only starburst regions, and
lack an underlying old stellar population.
In our sample, we include all BCDs as long as their outer isophotes
can be approximated as ellipses; our sample thus contains only
iE, nE, and iI galaxies.  
However, we note that our sample represents the whole BCD 
population well because BCDs of type iO are very rare.

All steps of the data reduction and analysis were performed 
using a standard CCD reduction process with the 
Image Reduction and Analysis Facilities (IRAF).
Raw images were first corrected for zero level bias and 
overscan pattern.
Each image was then divided by a flat field
obtained from the median of 5 -- 10 twilight and dome flat images.
During this process, pixels in the flat field image which
are more than 1.5 times the median or less than 0.5 times
the median are replaced with the median value.
In addition, we combined images of individual galaxies 
if multiple images taken at the same bands are available.
By combining images, we could not only correct for
cosmic rays, but also increase signal-to-noise ratios.
In the next stage, the sky brightness was subtracted from 
individual images.
For sky subtraction of most BCD images, we used separate sky images 
taken just after the observation of each galaxy.
However, if no sky image is available, we used the sky value
in the vicinity of a galaxy within the same image.
Finally, the processed images are trimmed centered at the galaxies.
Since most BCDs in our sample have angular sizes less than $2'$, 
the typical trimmed images have sizes of 
$\lesssim 500\times 500\ {\rm pixel}^2$.
If stars are located on or near the galaxies, we remove them by 
replacing them with the median value of the area near the stars.

\section{Determination of Apparent Axis Ratios}

To determine intrinsic axis ratio distribution, one
should start by measuring the apparent axis ratio, $q=1-\epsilon=
b/a$, of galaxies, as projected onto the sky.
Here $a$ and $b$ are the semimajor and semiminor
axes of a projected galaxy, and $\epsilon$ is the ellipticity.
From the obtained image of each BCD, 
we measure the ellipticities as a function of semimajor axis from the
center, using the Space Telescope Science Data Analysis System (STSDAS)
routine `isophote'.
In the routine, each galaxy was modeled as a nested series of ellipses,
yielding the semimajor axis, $a$, position angle, $\theta_{\rm pos}$, 
surface brightness, $\Sigma (a)$, and the positions of each ellipse's 
center.

The ellipticities of the isophotes of many 
stellar systems vary as a function of semimajor axis.
For BCDs we have the same problem of varying ellipticities, 
especially in the inner regions of BCDs.
The main reason that makes the isophotes of a BCD deviate from
well-aligned concentric ellipses is the existence of 
inner substructures, mainly bright HII regions 
(Thuan \& Martin 1981; Gordon \& Gottesman 1981).\markcite{
thuan1981, gordon 1981}
In Figure 1 we present, as an example, the $B$-band isophotes of ESO 495-G21, 
upon which are superimposed the best-fitting ellipses.
To better show the poor ellipse fitting in the inner region of the galaxy,
we expand the part enclosed by a box and present it in the  upper right
panel, in which the isophotes and corresponding best-fitting 
ellipses at $a=14.5$ and 16.5  are shown.
We also present the ellipticity 
as a function of the semimajor axis in the lower panel.
(Note that all length scales are in units of the pixel width.)
The isophotes are distorted by a bright HII region 
located to the right of the galaxy's center.
As a result, the obtained best-fitting ellipses are more elongated 
than those which would be obtained without the HII region,
and the position angle is tilted toward the direction of the HII region.
Beyond a certain point
($a \sim 30\ {\rm pixel}$ for ESO 495-G21) there no longer
exists any substructure, and the values of the ellipticity 
become stable as shown in the figure.
Note that the concentric circles centered at $(x,y) \sim (265,220)$
and $(x,y) \sim (262,189)$ are foreground stars, before they are removed 
for ellipse fitting.
The ellipse-fitting process fails as the surface brightness 
of a galaxy approaches background value;
for the case of ESO 495-G21 this point corresponds to $a\sim 95$.
Therefore, as a representative axis ratio of each BCD
we determine the intensity-weighted axis ratio averaged over the 
intermediate region where stable ellipticities can be obtained
with successful ellipse-fitting processes.
The intensity-weighted mean axis ratio is computed by 
$$
\bar{q} = {\int q(a) dL\over \int dL};\qquad
dL = 2\pi q a \left[ 1 + {1\over 2}{d {\rm ln}q(a)\over d{\rm ln}a}
\right] \Sigma (a) da,
\eqno(1)
$$ 
following the method by which Ryden \& Terndrup (1994)\markcite{ryden1994}
computed the axis ratios of dEs.
If more than a single band images are available, we take the mean
value of the ellipticities determined in individual bands
as a representative ellipticity.
The finally determined ellipticities of our sample galaxies 
are listed in Table 1.
We estimate the errors of $\bar{\epsilon}=1-\bar{q}$ by computing the 
variance of $\epsilon (a)$ within the range of semimajor axis where
ellipticities are measured.

A BCD with lots of star formation might have a $B$ image different 
from its $I$ image, for instance; the $B$ image emphasizes 
star formation regions, and hence is much clumpier. 
Thus, the measured ellipticities at $B$ might be systematically different 
from the ellipticities in other bands. To test this possibility,
in Figure 2 we compare the ellipticities measured in 
$B$ band, $\epsilon_{B}$, to those measured in $I$ band, $\epsilon_{I}$.
Ellipticities do not vary significantly
from one band to another. 
The values of ellipticities are consistent regardless of the observed 
bands because the determination of $\bar{\epsilon}$ depends strongly on
the ellipticities beyond the inner clumpy regions.

We also check the consistency of our measurements by measuring 
ellipticities in a different way.
In the second method, we measure the ellipticity $\epsilon_{24}$
at the point where the surface brightness is
$\Sigma_B = 24\ {\rm mag\ arcsec}^{-2}$.
For these determinations, we carried out $B$-band surface photometry 
of arbitrarily selected 16 sample BCDs by using elliptical isophote 
fitting method with the STSDAS routine `isophote'.
Photometric calibrations were carried out by using $\sim 50$ standard 
stars in the list of Graham (1982) and Landolt (1992).\markcite{
graham1982, landolt 1992} 
In Figure 3, we plot the correlation between $\bar{\epsilon}$ and
$\epsilon_{24}$.
The ellipticities determined in both ways 
are consistent each other.
In Figure 3, the error estimates of $\epsilon_{24}$ are derived
by computing the standard deviation of isophotes from their
corresponding best-fitting ellipses, i.e., 
$\sigma_{\epsilon_{24}} = \left[ \sum ({\bf x}_{\rm isophote}-
{\bf x}_{\rm ellipse})^2/n_{\rm point} \right]^{1/2}$. 
Here ${\bf x}_{\rm isophote}$ and ${\bf x}_{\rm ellipse}$ 
represent the positions of isophotes and ellipses on CCD pixel
coordinates, and $n_{\rm point}$ is the number of pixels on which
individual curves (both isophotes and ellipses) are drawn.

Since we use the same method for determining $\bar{\epsilon}$
as Ryden \& Terndrup (1994),\markcite{ryden1994}
we can directly compare our results for the apparent shapes
of BCDs to their results for the apparent shapes of dEs
in the Virgo Cluster. To the data obtained for 70 dEs by
Ryden \& Terndrup (1994), we add data for an additional
10 dEs observed in Virgo by Ryden et al. (1997).\markcite{ryden1997}
The cumulative distribution function of $\bar{q} = 1 - \bar{\epsilon}$
for the 62 BCDs is shown as the solid step function in Figure 4.
The cumulative distribution function of $\bar{q}$ for the
80 Virgo dEs is shown as the dotted step function in the same
Figure. A Kolmogorov-Smirnov test comparing the two distributions
shows a marginal difference, with $P_{\rm KS} = 0.057$. The main
difference between the apparent axis ratios of the two distributions
is that BCDs are slightly flatter, on average. For the
sample of BCDs, the mean and standard deviation of $\bar{q}$ are
$0.671 \pm 0.150$; for the sample of dEs, they are $0.719 \pm 0.156$.
A Mann-Whitney U-test comparing the medians of the two data sets
yields a probability $P_{\rm MW} = 0.068$. Thus, we can state
from the Mann-Whitney test that the two distributions differ at the 93.2\%
confidence level.

\section{Intrinsic Shape Determination}

The apparent axis ratio distributions of stellar systems provide 
important constraints about the 3-dimensional shapes of the systems
(Fasano \& Vio 1991; Lambas, Maddox, \& Loveday 1992; 
Tremblay \& Merritt 1995; Ryden 1992, 1996).\markcite{fasano1991, 
lambas1992, tremblay1995, ryden1992, ryden1996}
For the determination of intrinsic shapes, two methods are often 
applied. Both methods assume that the galaxies which we
observe are randomly oriented with respect to us.
The first method for determining intrinsic shapes
assumes that the galaxies are intrinsically axisymmetric,
and that they are either all oblate or all prolate.
With these assumptions, one can perform a unique mathematical
inversion to go from the distribution $f(q)$ of apparent shapes
to the distribution $f(\gamma)$ of intrinsic shapes.
The second method assumes that galaxies are triaxial; in
this case one no longer has a unique solution to the problem.
However, if one assumes a functional form (e.g., Gaussian) for
the distribution of intrinsic axis ratios, one can vary the
parameters of the function until one finds the model which
best fits the observed distribution of apparent axis ratios.
Parametric fits are useful because they enable us to see how statistics
such as the $\chi^2$ and KS scores vary as the parameters are changed.
We estimate the intrinsic shapes of BCDs by using both the non-parametric and
the parametric methods. 

\subsection{Non-parametric Method}

Using the non-parametric method, we reject or accept, at 
a known confidence level, the null hypothesis that the intrinsic shapes 
of the galaxies in a sample are randomly oriented oblate
spheroids. Moreover, if
the null hypothesis is not rejected, the non-parametric
method gives an estimate for the distribution function $f_{\rm o}$
of the galaxies' intrinsic axis ratios, $\gamma \leq 1$.
Similarly, we can reject or accept the hypothesis that 
they are randomly oriented prolate spheroids.
To accomplish this, we make a non-parametric kernel
estimate $f(q)$ of the distribution of the apparent
axis ratio $q$.
We then mathematically invert 
$f(q)$ to find $f_{\rm o}(\gamma )$ and $f_{\rm p}(\gamma )$,
the estimated distributions of the intrinsic axis ratio $\gamma$ for
a population of oblate and prolate spheroids, respectively.
Confidence intervals are placed on the estimators
$f(q)$, $f_{\rm o}$, and $f_{\rm p}$ by performing repeated
bootstrap resampling of the observed data and creating new 
estimates from each sample.
The spread in the bootstrap estimates of $f$ at a given value of
$q$ and of $f_{\rm o}$ and $f_{\rm p}$ at a given value of $\gamma$
provides confidence intervals for the non-parametric 
estimates of these functions.
Once the estimators $f_{\rm o}$ and $f_{\rm p}$ are determined,
one can exclude the oblate or prolate hypothesis at a given
confidence level
if the upper confidence limit drops below zero for 
any value of the intrinsic axis ratio $\gamma$.
For the details of the non-parametric kernel estimators, see
Ryden (1996) and Tremblay \& Merritt (1995).\markcite{ryden1996,tremblay}

In the upper panel of Figure 5a, we present the 
non-parametric kernel estimate of the distribution of the 
apparent axis ratios for our sample of 62 BCDs.
In the following panels, we also present the distributions of intrinsic
axis ratios assuming that the BCDs are oblate (middle) and
prolate (bottom).
The solid line in each panel is the best estimate, the dashed 
lines are the 80\% confidence band, and the dotted lines are the 98\%
confidence band. (That is, 10\% of the bootstrap resamplings lie
above the 80\% confidence band, and 10\% lie below it.
Similarly, 1\% of the resamplings lie above the 98\% confidence
band, and 1\% lie below it.)
The hypothesis that BCDs are randomly oriented oblate spheroids
is ruled out at the 99\% (one-sided) confidence level. Due to the lack of
nearly circular BCDs (those with $q \sim 1$), the 98\% confidence
band for $f_{\rm o}$ dips far below zero at $\gamma > 0.9$.
On the other hand, the prolate hypothesis cannot be ruled
out at the 99\% (one-sided) confidence level. The best-fitting
prolate distribution (the solid line in the
bottom panel of Figure 5a) gives a mean and standard deviation for
the intrinsic axis ratio $\gamma$ of $0.58 \pm 0.13$.

For the comparison of intrinsic shape of BCDs with that of dEs
we also present $f(q)$, $f_{\rm o}$, and $f_{\rm p}$ for 
the sample of 80 Virgo dEs, in Figure 5b.
We find that the oblate hypothesis cannot be ruled out at the 99\%
(one-sided) confidence level for this sample,
while the prolate hypothesis is 
{\it perfectly} acceptable. The best-fitting prolate distribution
gives a mean and standard deviation for $\gamma$ of $0.63 \pm 0.15$.
Ryden (1996)\markcite{ryden1996},
analyzing a set of 170 Virgo dEs observed by 
Binggeli \& Cameron (1993),\markcite{binggeli1993}
found that for this larger sample,
the oblate hypothesis can be ruled out at the 99\% confidence
level. By contrast, Ichikawa (1989),\markcite{ichikawa1989}
analyzing a sample of 98 Virgo dEs, alleges that their
apparent shapes are consistent with their being a population
of oblate spheroids. 
Instead of doing a non-parametric analysis, Ichikawa used Gaussian
approximations to the correct values of
$f_{\rm o}$ and $f_{\rm p}$, finding that a Gaussian
with mean and standard 
deviation $(\gamma_0, \sigma)\sim (0.6,0.06)$ is the best fit 
for a population of oblate spheroids.
However, although this is the best fit, it is not a good fit.
Although he does not provide a $\chi^2$ probability, fig 1(a)
of his paper shows very poor fit; the best-fitting distribution 
overestimates the number of nearly circular dEs and severely underestimates
the number of apparently highly flattened dEs. Thus, Ichikawa's
results, like ours, indicate that the oblate hypothesis yields
a poor fit to the observed distribution of apparent axis ratios.
Ichikawa also presented the best-fitting distribution under prolate 
hypothesis with $(\gamma_0, \sigma)\sim (0.64,0.12)$, 
which is very similar to our determination of the mean and
standard deviation of $\gamma$ for prolate dEs.

\subsection{Parametric Method}

For the parametric determination of the intrinsic axis ratio
distribution of BCDs, we follow the assumptions of Ryden \& Terndrup 
(1994)\markcite{ryden1994}; that galaxies are 
triaxial ellipsoids with axis ratios $1 \geq  \beta \geq \gamma$, and that
the distribution of intrinsic axis ratios follows a Gaussian 
distribution with means $\beta_0$ and $\gamma_0$ and a common
width $\sigma$; i.e., 
$$
f(\beta,\gamma )  \propto
\exp \left[ -
{(\beta-\beta_0)^2 + (\gamma-\gamma_0)^2
\over
2\sigma^2
}
\right].
\eqno(2)
$$
With these assumptions we produce a large number ($10^4$) of
test BCDs whose intrinsic axis ratios are distributed according to
equation (2).
Then we compute their projected axis ratios assuming that BCDs are 
randomly oriented with respect to us.
When an triaxial ellipsoid is projected with the viewing angles of
$\theta$ and $\phi$, it appears as an ellipse with an
apparent axis ratio of
$$
q(\beta,\gamma,\theta,\phi) =
\left[ {
A + C - \sqrt{(A-C)^2+B^2}
\over 
A + C + \sqrt{(A-C)^2+B^2}
} \right]^{1/2},
\eqno(3)
$$
where
$$
\cases{
A = (\cos^2 \phi + \beta^2\sin^2\phi)\cos^2\theta + \gamma^2\sin^2\theta,\cr
B = \cos\theta\sin 2\phi (1-\beta^2),\cr
C = \sin^2\phi + \beta^2\cos^2\phi 
}
\eqno(4)
$$ 
(Binney 1985).
The cumulative distribution of the projected axis ratios of the 
test BCDs is then compared with the observed one by using 
Kolmogorov-Smirnov (KS) tests;
we additionally apply $\chi^2$ tests to the binned distributions.

Figure 6a and 6b show the isoprobability contours on 4 slices 
through the $(\beta_0,\gamma_0,\sigma)$ parameter space,
as measured by KS and $\chi^2$ tests, respectively.
When measured by KS tests the best-fitting distribution has parameters of
$(\beta_0,\gamma_0,\sigma) = (0.66, 0.55, 0.16)$ with the KS 
probability of $P_{\rm KS} = 0.99$, implying that the intrinsic
shape of BCDs can be well fitted by a
population of triaxial ellipsoids.
We obtain consistent results when measured by $\chi^2$ tests;
the best-fitting distribution has parameters of 
$(\beta_0,\gamma_0,\sigma) = (0.77, 0.51, 0.16)$ with 
the $\chi^2$ probability of $P_{\chi^2} = 0.96$.

For the 80 dE galaxies from Ryden \& Terndrup (1994) and
Ryden et al. (1997), the best-fitting parametric model,
as measured by a KS test, had $(\beta_0,\gamma_0,\sigma) =
(0.85,0.64,0.24)$, with $P_{\rm KS} = 0.97$. In Figure 7,
we show four slices through parameter space, showing the
isoprobability contours for the parametric dE models as measured
by KS tests. When the best-fitting distributions for dEs
are compared to that of BCDs, one finds that the dEs are
rounder, but the difference is not greatly significant.

In Figure 4, we present the cumulative distributions of 
apparent axis ratios for BCDs 
(solid step function) and dEs (dotted step function).
On top of each observed distribution we overlay the
expected distribution from the best-fitting triaxial model,
as measured by a KS test.
For BCDs, the prediction of the best-fitting model, with
$(\beta_0,\gamma_0,\sigma) = (0.66, 0.55, 0.16)$, is
given by the solid smooth line.
For dEs, the prediction of the best-fitting model, with
$(\beta_0,\gamma_0,\sigma) = (0.85,0.64,0.24)$, is
given by the dotted smooth line.

\section{Discussion}

The results of our morphological analysis based on 62 BCDs and 80 dEs
are summarized as follows.

\begin{minipage}{6.3in}
\begin{flushleft}
1.\ The apparent shapes of both BCDs and dEs can be explained
by their being either triaxial ellipsoids or prolate spheroids.

2.\ However, it is unlikely that the intrinsic shapes of either
BCDs and dEs are purely oblate spheroids.

3.\ Compared to dEs, BCDs are flatter,
but the difference between the apparent axis ratio distributions of 
BCDs and dEs is marginal.  

\end{flushleft}
\end{minipage}

The nature of BCDs remains ambiguous and confusing
(Papaderos et al.\ 1996)\markcite{papaderos1996}, and thus  
theories to explain their nature are very controversial.
There are two major competing hypotheses.
The first one claims that BCDs are basically a different population from dEs.
According to this scenario BCDs are truly young systems, in which the
present starburst is the first in the galaxy's lifetime.
The other model suggests that BCDs might be dEs which are mainly
composed of old stellar populations, and that the observed
spectroscopic features and the spectral energy distributions 
are attributed to a recent burst of star formation.
That is, BCDs are basically dEs seen during a brief episode
($t \sim 10^7\ {\rm yr})$ of violent star formation.

Although strong conclusions cannot be derived just from
a single comparison of morphology, the similarity
between the intrinsic shapes of BCDs and dEs may
give support to the second hypothesis that BCDs are
just a sub-population of dEs that have undergone a recent
burst of star formation.
The second hypothesis is also supported by other observations.
Thuan (1983)\markcite{thuan1983}
argued that the near-infrared 
emission in the vast majority of BCDs is attributable to old K and M stars, 
which are the major components of dEs;
Hunter \& Gallagher (1985)\markcite{hunter1985} also agreed that 
young stars are not the major contributors
to the near-infrared flux of BCDs.
In addition, many BCDs, like dEs, have surface brightness profiles 
that are well fitted by exponentials.
For example, Haro 2 was classified as a BCD by Haro
(1956)\markcite{haro1956} and Markarian (1967)\markcite{markarian1967},
based on its spectral features.
By contrast, Loose \& Thuan (1986)\markcite{loose1986} claimed 
that Haro 2 should be classified as a dE, based on its surface 
brightness profiles.

The question of how to explain the difference in spectra
between the two types of dwarf galaxy still remains.
Drinkwater \& Hardy (1991)\markcite{drinkwater} stated
that BCDs, after the end of starburst activity, fade
to form dEs, as the amount of gas decreases by stripping.
On the other hand, Silk, Wyse, \& Shields (1987)\markcite{silk1987}
proposed a totally reversed evolution sequence: from dEs to BCDs.
According to them, hot gas, left over from the galaxy formation era,
is trapped in a galaxy group, cools and accretes onto dEs, and provokes
an intense burst of star formation.
As a result, dEs are transformed into dEs.
Although our results cannot clearly designate the direction of
evolution, they suggest some evolutionary relation between the
two systems.

\acknowledgements
We would like to thank staffs  at the Mt.\ Stromlo \&
Siding Spring Observatory,
especially K.\ C.\ Freeman for supporting observations.
We also thank D.\ M.\ Terndrup, R.\ W.\ Pogge, and T.\ R.\ 
Lauer for kindly providing additional flattening data of dwarf 
ellipticals.
E.-C.\ S.\ has been supported by Basic Research Fund of Korea 
Astronomy Observatory.
B.\ S.\ R.\ was supported by grant NSF AST-93-577396.

%REFERENCES

\clearpage

%**************** table

\begin{center}
\bigskip
\bigskip
\centerline{\small {TABLE 1}}
\smallskip
\centerline{\small {\sc The Intensity Weighted Mean Ellipticities}}
\smallskip
\begin{tabular}{lll|lll}
\hline
\hline
\multicolumn{1}{c}{name} &
\multicolumn{1}{c}{band} &
\multicolumn{1}{c}{$\bar{\epsilon} \pm \Delta\epsilon$} &
\multicolumn{1}{|c}{name} &
\multicolumn{1}{c}{band} &
\multicolumn{1}{c}{$\bar{\epsilon} \pm \Delta\epsilon$} \\
\hline
CTS 1010   & $V$, $R$             &0.387 $\pm$ 0.064 &  ESO 508-G33  & $B$                & 0.149 $\pm$ 0.202  \\
CTS 1020   & $B$, $I$             &0.244 $\pm$ 0.021 &  ESO 554-IG27 & $B$, $V$, $R$, $I$ & 0.323 $\pm$ 0.057  \\
CTS 1025   & $V$                  &0.352 $\pm$ 0.070 &  Fairall 67   & $B$, $V$, $R$, $I$ & 0.371 $\pm$ 0.010  \\
CTS 1033   & $B$                  &0.577 $\pm$ 0.000 &  Fairall 177  & $B$, $V$, $R$, $I$ & 0.234 $\pm$ 0.038  \\
CTS 1034   & $B$                  &0.089 $\pm$ 0.207 &  Fairall 318  & $B$, $V$           & 0.405 $\pm$ 0.025  \\
CTS 1037   & $B$, $V$, $R$        &0.141 $\pm$ 0.007 &  Fairall 346  & $B$, $V$, $R$, $I$ & 0.188 $\pm$ 0.011  \\
CTS 1040   & $B$, $V$, $R$        &0.205 $\pm$ 0.020 &  G0003-41     &                $I$ & 0.463 $\pm$ 0.037  \\
CTS 1042   & $B$, $V$, $R$, $I$   &0.357 $\pm$ 0.010 &  Haro 6       & $B$, $V$, $R$, $I$ & 0.136 $\pm$ 0.021  \\
ESO 037-G03 & $B$, $V$, $I$       &0.411 $\pm$ 0.028 &  Haro 14      & $B$, $V$, $R$, $I$ & 0.139 $\pm$ 0.025  \\
ESO 092-G02 & $V$                 &0.331 $\pm$ 0.056 &  Haro 15      & $B$, $V$, $R$, $I$ & 0.259 $\pm$ 0.035  \\
ESO 102-G14 & $B$                 &0.317 $\pm$ 0.015 &  Haro 18      & $B$, $V$, $R$, $I$ & 0.143 $\pm$ 0.022  \\
ESO 105-IG11& $B$, $V$, $R$, $I$  &0.458 $\pm$ 0.031 &  Haro 20      & $B$, $V$, $R$, $I$ & 0.474 $\pm$ 0.008  \\
ESO 148-IG07& $B$, $R$            &0.200 $\pm$ 0.059 &  Haro 21      & $B$, $V$, $R$, $I$ & 0.211 $\pm$ 0.037  \\
ESO 156-G38 & $B$, $V$, $R$, $I$  &0.327 $\pm$ 0.025 &  II Zw 40     & $V$                & 0.415 $\pm$ 0.035  \\
ESO 185-IG13& $B$, $V$, $R$, $I$  &0.180 $\pm$ 0.025 &  Mrk 49       & $B$, $V$, $R$, $I$ & 0.226 $\pm$ 0.012  \\
ESO 249-G13 & $B$, $V$            &0.686 $\pm$ 0.046 &  Mrk 400      & $B$, $V$, $R$, $I$ & 0.481 $\pm$ 0.021  \\
ESO 250-G03 & $B$, $V$, $R$, $I$  &0.122 $\pm$ 0.020 &  Mrk 527      & $B$                & 0.364 $\pm$ 0.019  \\
ESO 286-IG19& $B$, $V$            &0.389 $\pm$ 0.046 &  NGC 1705     & $B$, $V$, $R$, $I$ & 0.274 $\pm$ 0.014  \\
ESO 338-IG04& $B$, $V$            &0.343 $\pm$ 0.016 &  NGC 5253     & $B$, $V$, $R$, $I$ & 0.527 $\pm$ 0.010  \\
ESO 352-G67 & $B$, $V$, $R$, $I$  &0.164 $\pm$ 0.023 &  Pox 36       & $B$, $V$, $R$, $I$ & 0.540 $\pm$ 0.024  \\
ESO 379-G17 & $B$, $V$, $I$       &0.080 $\pm$ 0.121 &  Pox 139      &                $I$ & 0.510 $\pm$ 0.080  \\ 
ESO 386-G19 & $B$, $V$, $R$, $I$  &0.144 $\pm$ 0.022 &  T1258-363    & $B$                & 0.340 $\pm$ 0.033  \\
ESO 422-G03 & $B$, $V$, $R$, $I$  &0.458 $\pm$ 0.006 &  T1324-276    & $B$, $V$, $R$, $I$ & 0.569 $\pm$ 0.011  \\
ESO 435-IG20& $B$, $V$, $R$, $I$  &0.412 $\pm$ 0.019 &  T2138-397    & $B$, $V$, $R$, $I$ & 0.341 $\pm$ 0.039  \\
ESO 462-IG20& $B$, $V$, $R$, $I$  &0.158 $\pm$ 0.030 &  T2259-398    & $B$, $V$,      $I$ & 0.041 $\pm$ 0.010  \\
ESO 480-IG08& $B$, $V$, $R$, $I$  &0.296 $\pm$ 0.009 &  T2311-411    & $B$                & 0.313 $\pm$ 0.033  \\
ESO 480-IG12& $B$, $V$, $I$       &0.545 $\pm$ 0.017 &  Tololo 3     & $B$, $V$, $R$, $I$ & 0.311 $\pm$ 0.042  \\
ESO 483-G13 & $B$, $V$, $R$, $I$  &0.394 $\pm$ 0.009 &  UM 40        & $B$, $V$, $R$, $I$ & 0.467 $\pm$ 0.022  \\
ESO 495-G21 & $B$, $V$, $R$, $I$  &0.158 $\pm$ 0.013 &  UM 69        & $B$, $V$, $R$, $I$ & 0.589 $\pm$ 0.010  \\
ESO 502-IG11& $V$                 &0.547 $\pm$ 0.041 &  UM 448       & $B$, $V$, $R$, $I$ & 0.311 $\pm$ 0.016  \\
ESO 505-G12 & $B$, $V$, $R$, $I$  &0.259 $\pm$ 0.014&  UM 621        & $B$, $V$, $R$      & 0.520 $\pm$ 0.017  \\
\hline
\end{tabular}
\end{center}
\smallskip
\noindent
{\footnotesize
\qquad NOTE.---
The intensity weighted mean ellipticities of 62 BCDs.
Also listed are the observed bands for each image.
}

\bigskip
\postscript{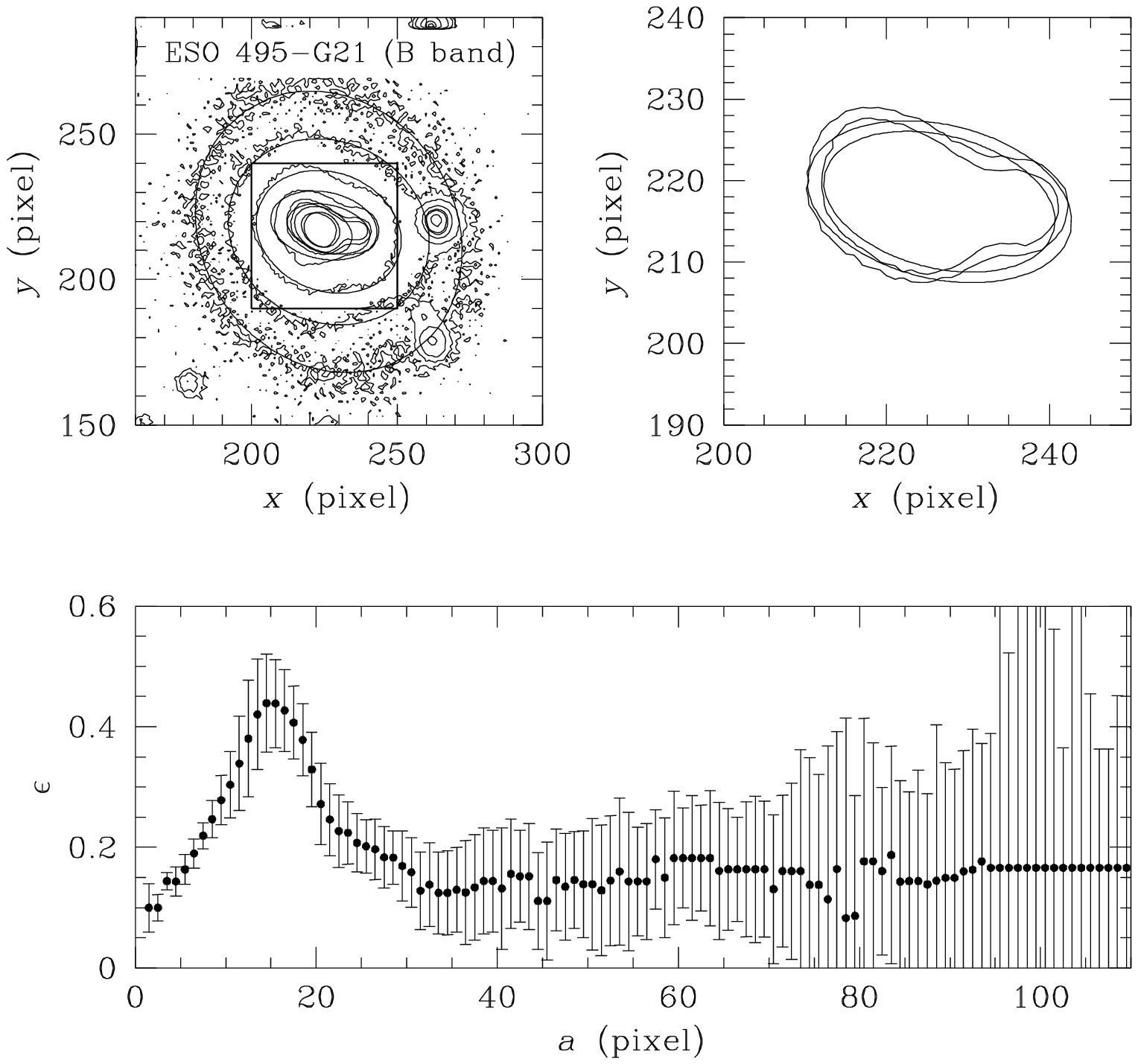}{0.70}
\noindent
{\footnotesize {\bf Figure 1:}\
Upper panels: the $B$-band isophotes of ESO 495-G21,
with the best-fitting ellipses superimposed.
To better show the ellipse fits for the inner isophotes, the 
region enclosed by a box is expanded and shown in the upper right
panel, where the best-fitting ellipses at $a=14.5$ and 16.5
and corresponding isophotes are presented.
Note the length scales are given by pixel values, 
which correspond to $0''\hskip-2pt .57/{\rm pixel}^{-1}$.
Lower panel: 
the determined ellipticity of ESO 495-G21 as a function of
semimajor axis.  
Error estimates for $\epsilon (a)$ are derived by computing the
standard deviation of isophotes from their best-fitting ellipses.
}
\clearpage

\bigskip
\postscript{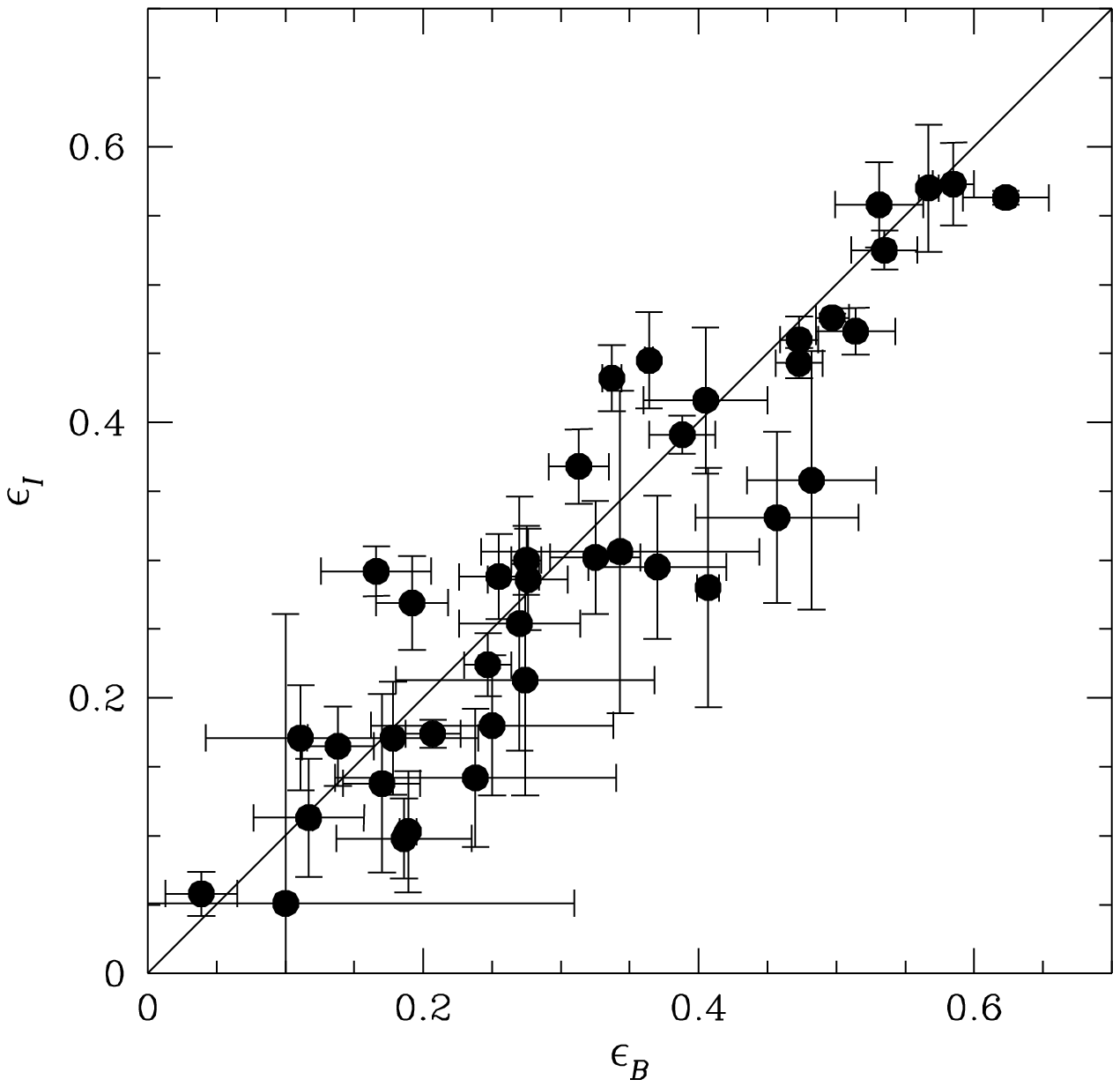}{0.60}
\noindent
{\footnotesize {\bf Figure 2:}\
The correlation between ellipticities determined in 
$B$ and $I$ bands.
The solid line across the panel represents the identity
between the two measurements, and is not a fit to the data.
The errors in each band are estimated by computing the
standard deviations of isophotes from the best-fitting ellipses.
}
\clearpage

\bigskip
\postscript{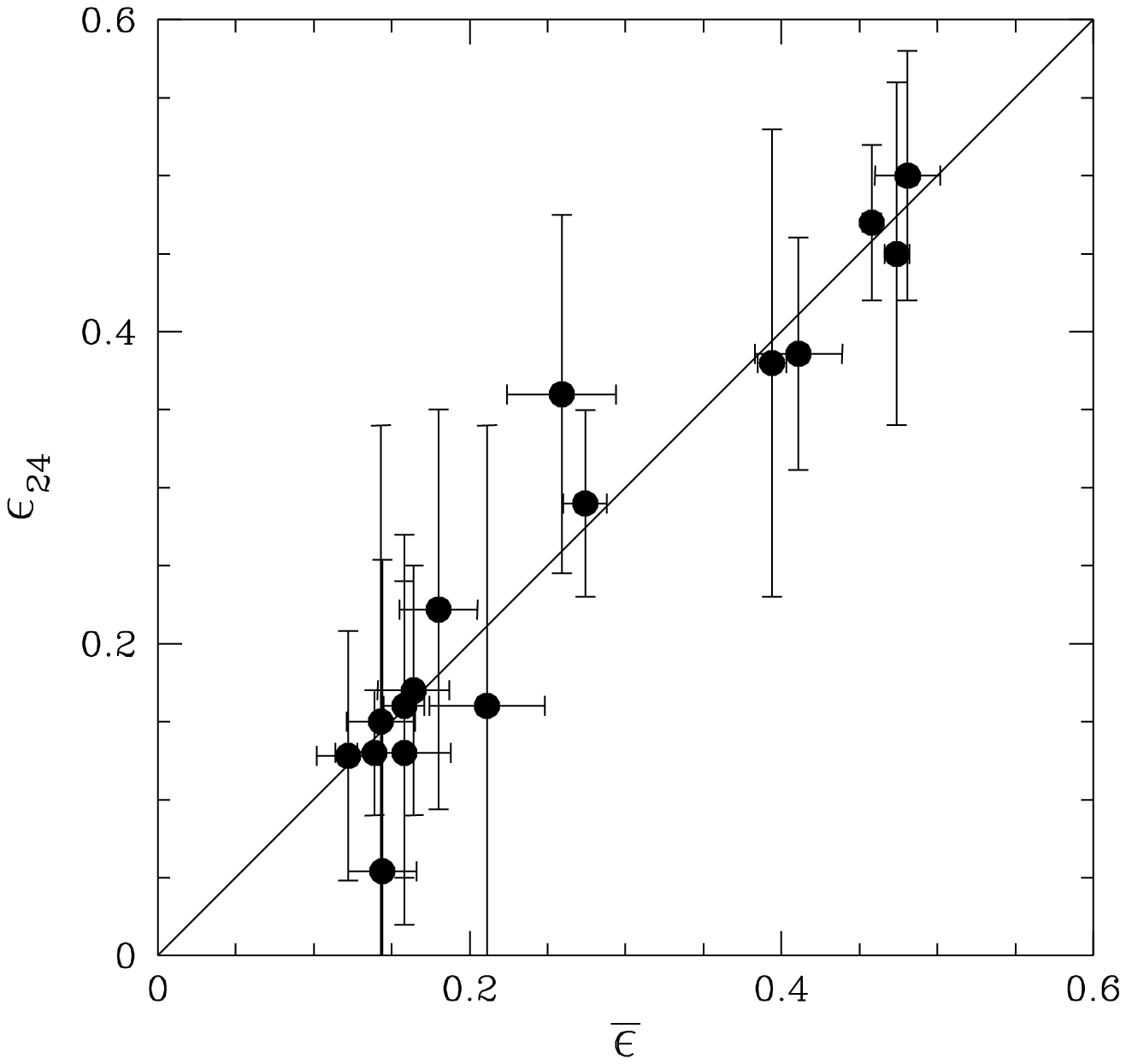}{0.60}
\noindent
{\footnotesize {\bf Figure 3:}\
The correlation between the intensity-weighted ellipticity,
$\bar{\epsilon}$, and the ellipticity of the
$\Sigma_{B} = 24$ mag isophote,
$\epsilon_{24}$, for a randomly selected sample of 16 BCDs.
The solid line across the panel represents the identity 
between the two measurements.
The errors of $\epsilon_{24}$ are estimated by computing the 
standard deviations of isophotes from the best-fitting ellipses, 
while those for $\bar{\epsilon}$ are determined from 
the variance of $\epsilon (a)$.
}
\clearpage

\bigskip
\postscript{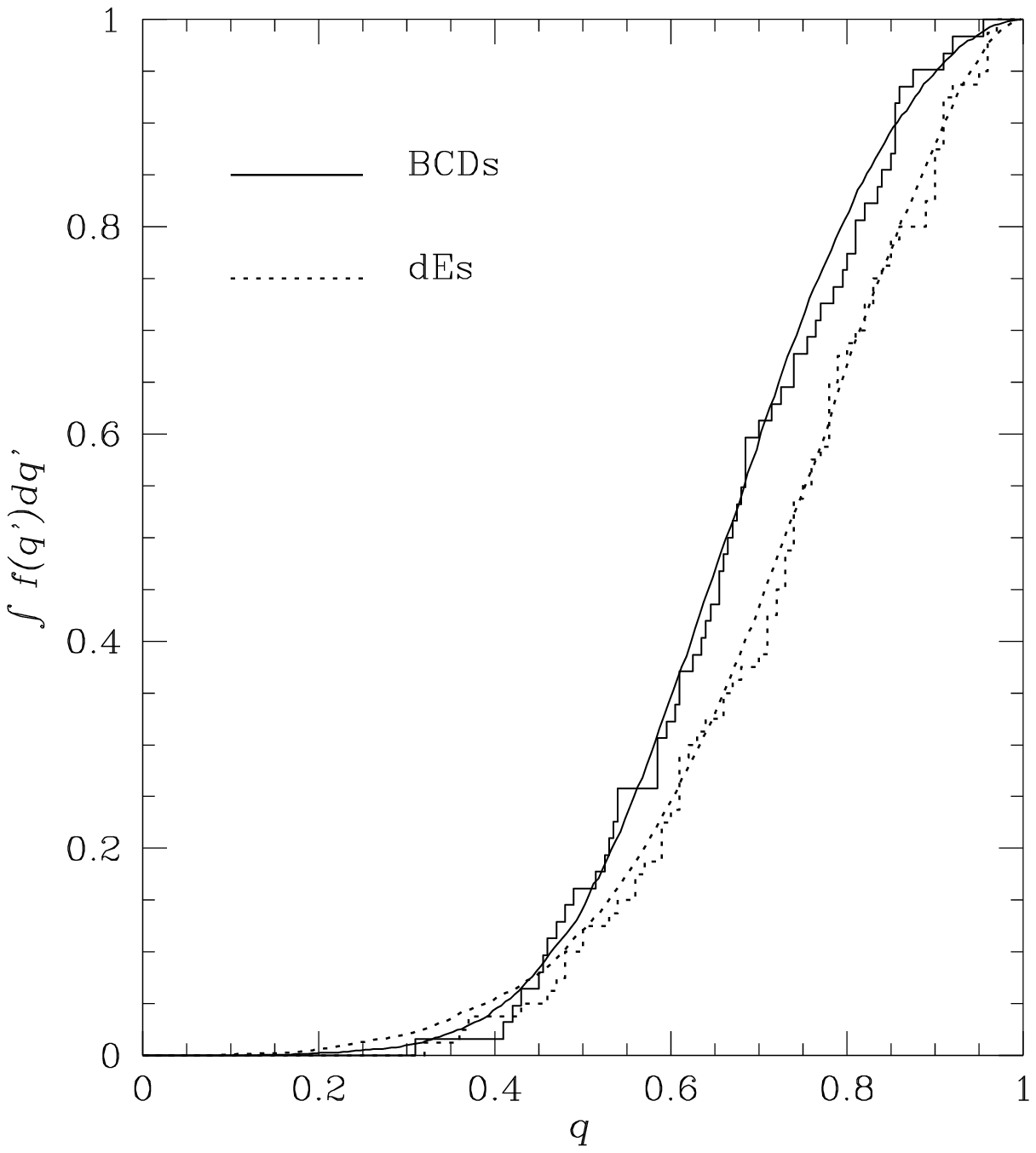}{0.68}
\noindent
{\footnotesize {\bf Figure 4:}\
The cumulative distribution of apparent axis ratios for
62 BCDs (solid line) and 80 dEs (dotted line). The step function,
in each case, is the data;
the smooth curve superimposed is the predicted
distribution of apparent shapes from the best-fitting
triaxial model.
}
\clearpage

\bigskip
\postscript{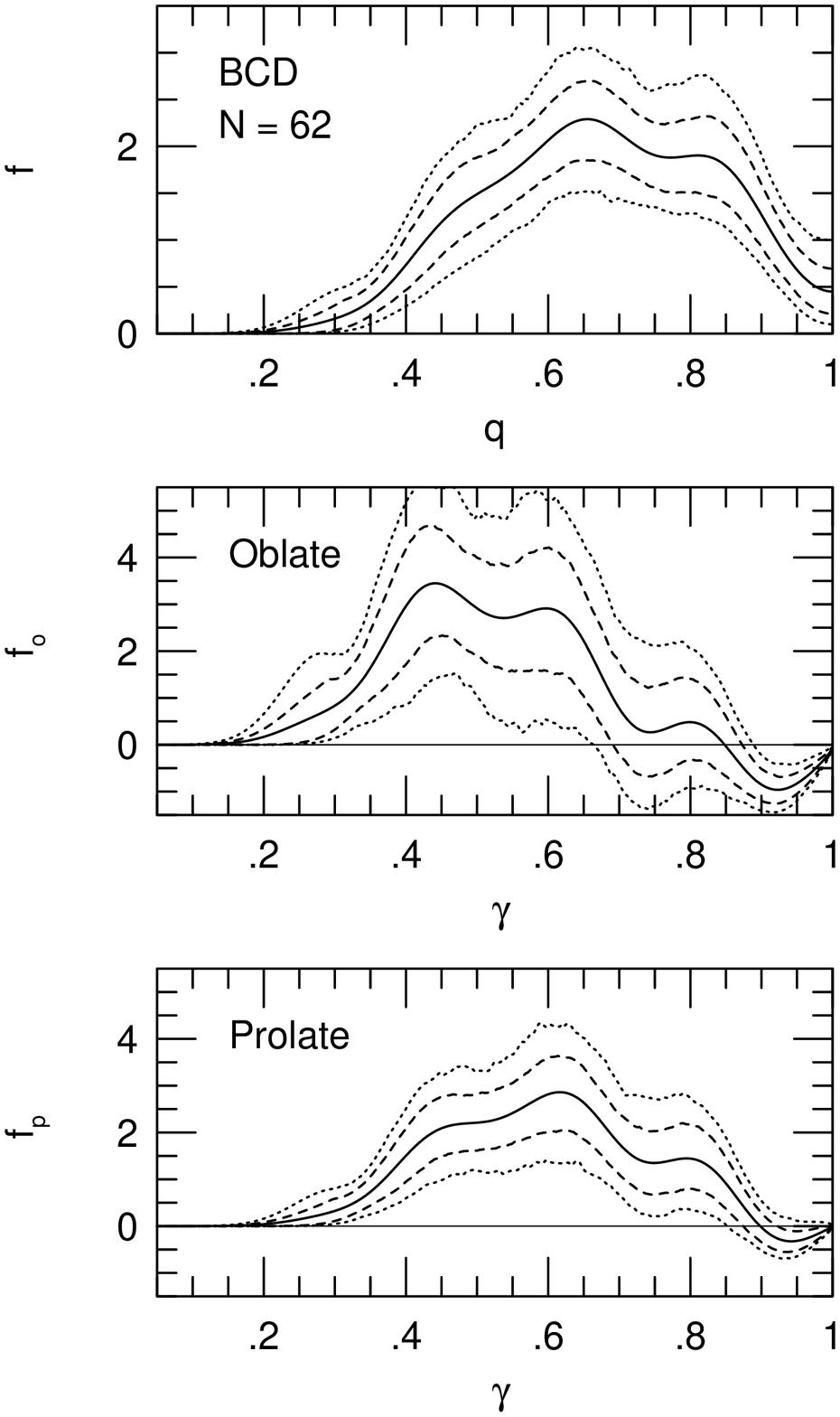}{0.85}
\noindent
{\footnotesize {\bf Figure 5a:}\
The non-parametric kernel estimate of the distribution of a sample
of 62 BCDs (top panel).
Also shown are the distributions of intrinsic axis ratios, 
assuming BCDs are all oblate (middle panel) and all prolate
(bottom panel).
The solid line in each panel is the best estimate, the dashed 
lines are the 80\% confidence band, and dotted lines are the
98\% confidence band.
}
\clearpage

\bigskip
\postscript{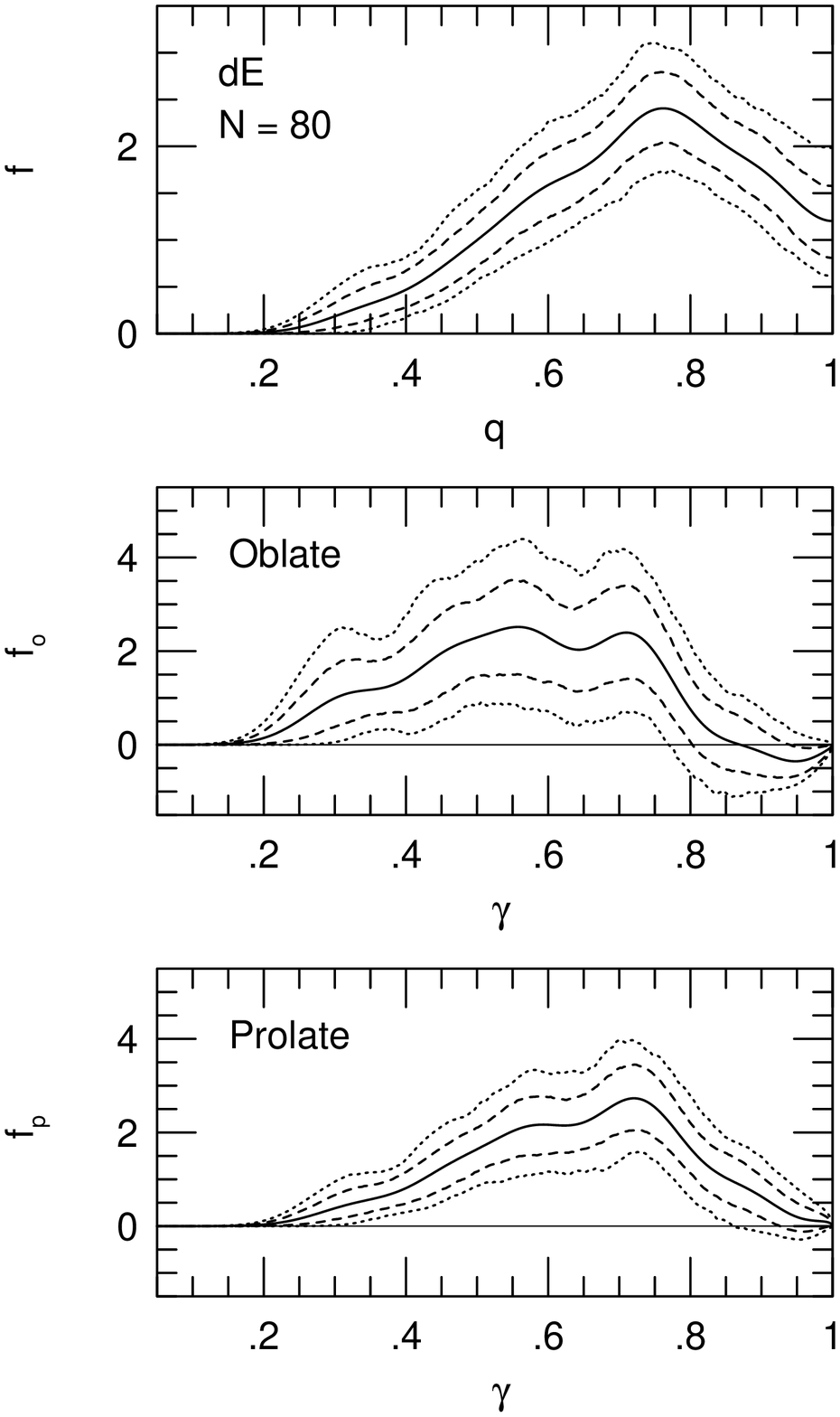}{0.85}
\noindent
{\footnotesize {\bf Figure 5b:}\
The non-parametric kernel estimate of the distribution of 
a sample of 80 dEs (top panel).
The distributions of intrinsic axis ratios under all oblate
(middle panel) and all prolate (bottom panel) are also shown.
The confidence bands are same as in Figure 5a.
}
\clearpage

\bigskip
\postscript{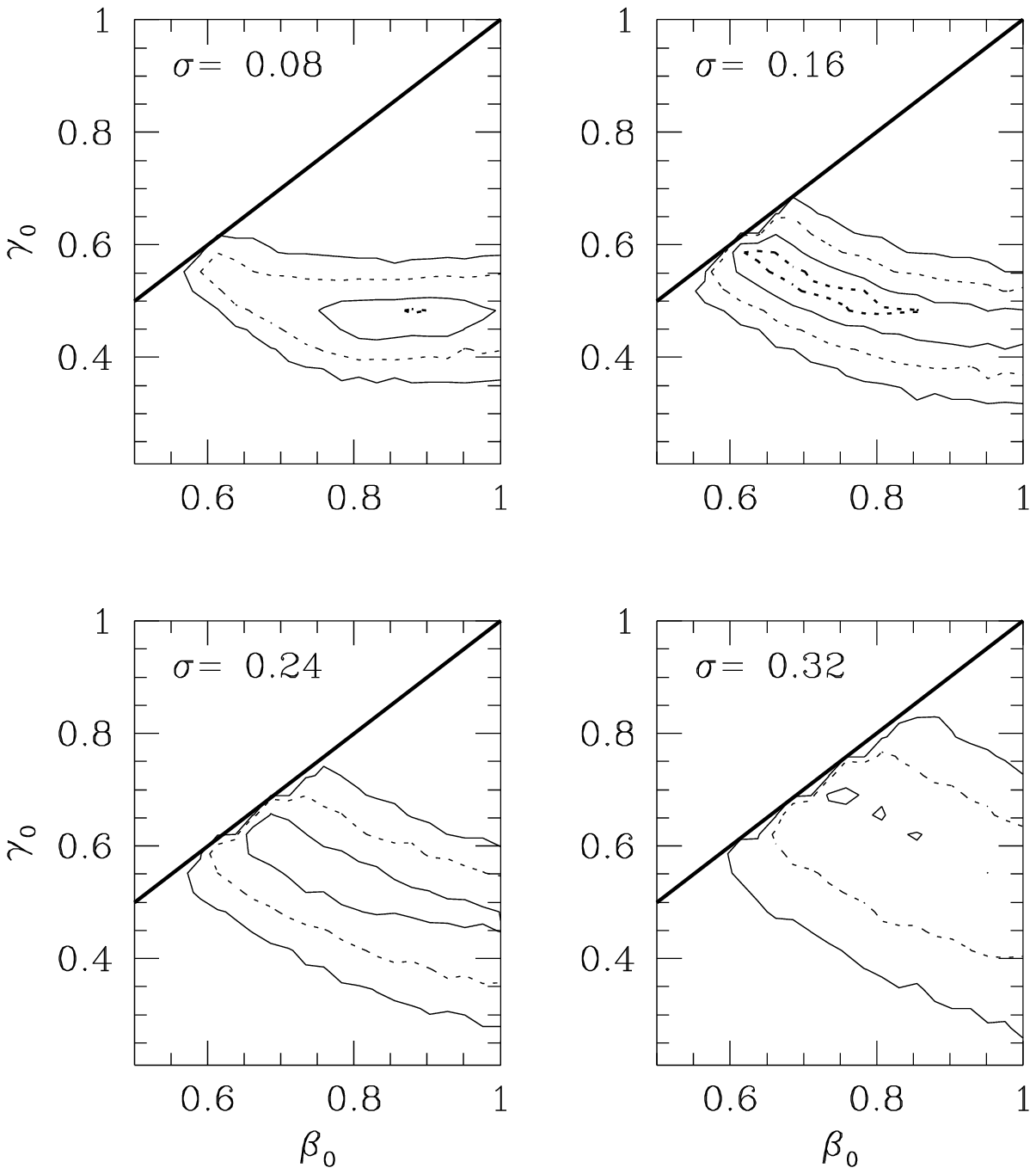}{0.68}
\noindent
{\footnotesize {\bf Figure 6a:}\
Isoprobability contours, as measured by KS tests, on 4 slices through
$(\beta_0,\gamma_0,\sigma)$ parameter space.
The fits are for the 62 galaxies in the BCD sample.
Contours are drawn at the levels $P_{\rm KS} = 0.01$,
0.1, 0.5, and 0.9 from outside.
}
\clearpage

\bigskip
\postscript{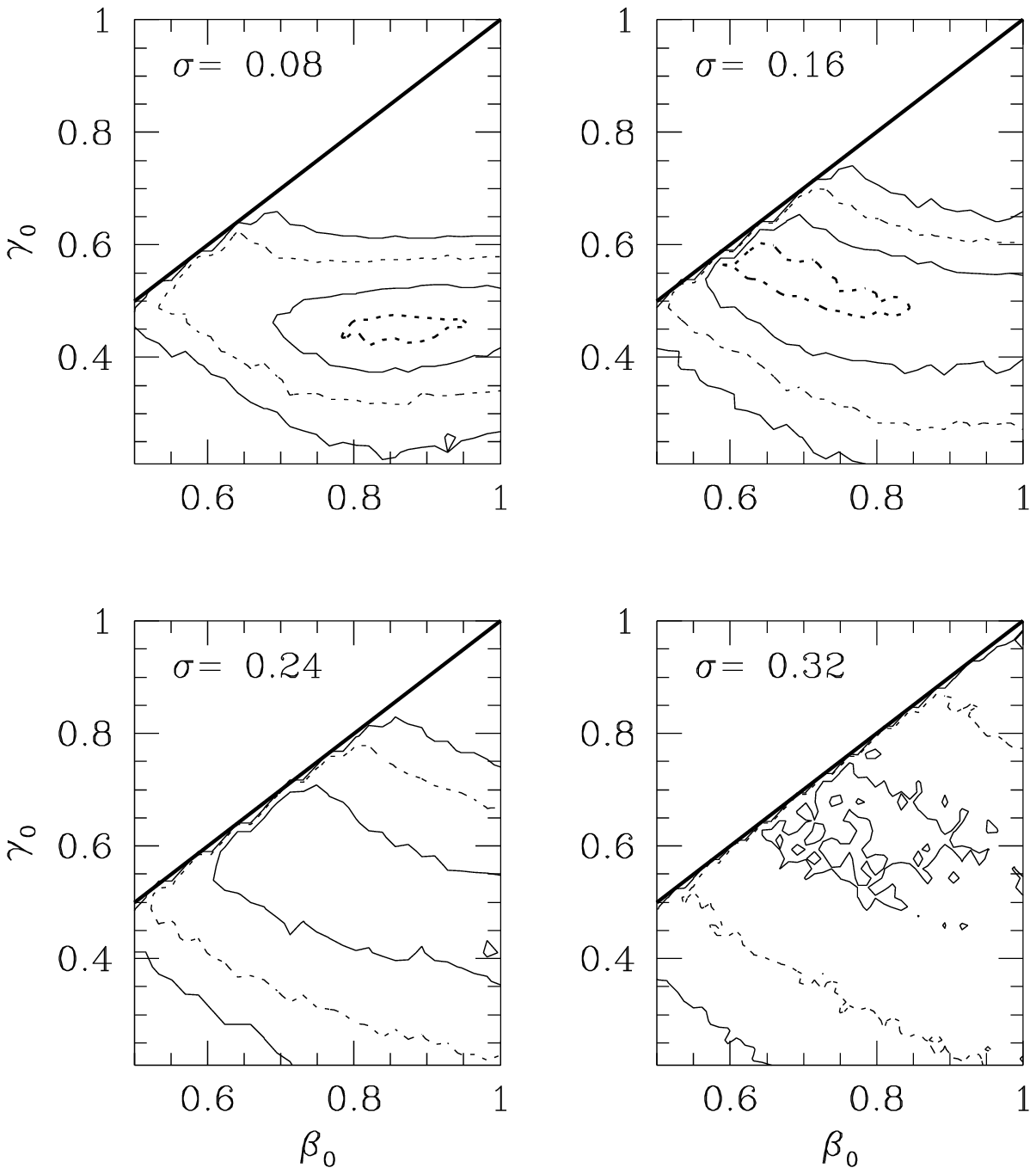}{0.68}
\noindent
{\footnotesize {\bf Figure 6b:}\
Isoprobability contours, as measured by $\chi^2$ tests, on 4 slices through
$(\beta_0,\gamma_0,\sigma)$ parameter space.
The fits are for the 62 galaxies in the BCD sample.
Contours are drawn at the levels $P_{\chi^2} = 0.01$, 0.1, 0.5,
and 0.9 from outside.
}
\clearpage

\bigskip
\postscript{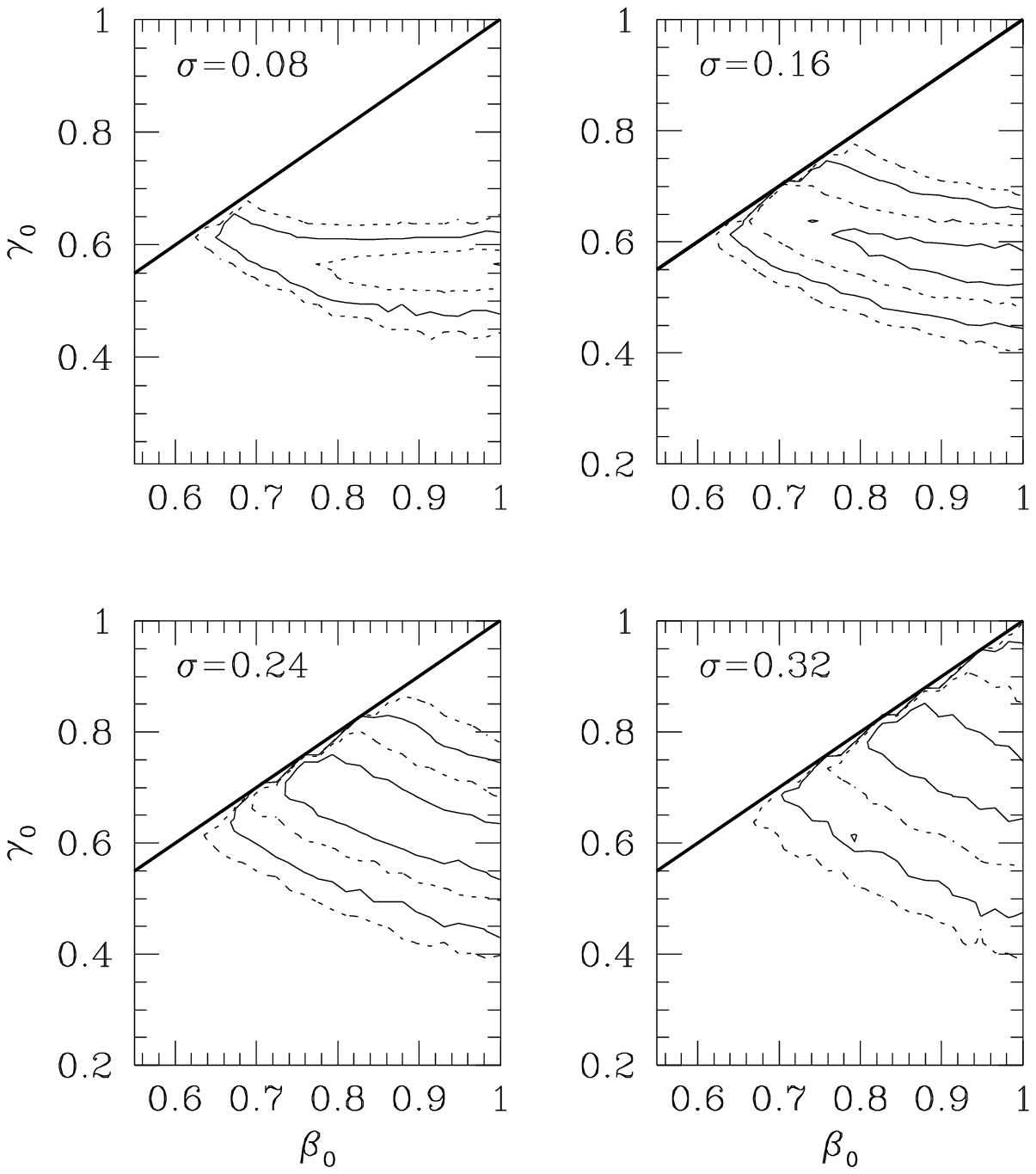}{0.68}
\noindent
{\footnotesize {\bf Figure 7:}\
Isoprobability contours, as measured by KS tests, on 4 slices through
$(\beta_0,\gamma_0,\sigma)$ parameter space. The fits
are for the 80 galaxies in the dE sample. Contours are drawn
at the levels $P_{\rm KS} = 0.01$, 0.1, 0.5, and 0.9.
}
\clearpage


\clearpage
\begin{references}
\reference{acker1991}         Acker, A., Stenholm, B., \& Verson, P.\ 1991,
			      \aaps, 87, 449
\reference{bergvall1986}      Bergvall, N., \& Olofsson, K.\ 1986,
			      \aaps, 64, 469
\reference{bessell1990}       Bessell, M.\ S.\ 1990, \pasp, 102, 1181
\reference{binggeli1985}      Binggeli, B., Sandage, A., \& Tammann, G.\ A.\
                              1985, \aj, 90, 1681
\reference{binggeli1993}      Binggeli, B., \& Cameron, L.\ M.\ 1993,
                              \aaps, 98, 297
\reference{binney1985}        Binney, J.\ 1985, \mnras, 212, 767
\reference{caldwell1983}      Caldwell, N.\ 1983, \aj, 88, 804
\reference{caldwell1987}      Caldwell, N., \& Bothun, G.\ D.\ 
		              1987, \aj, 94, 1126
\reference{drinkwater1991}    Drinkwater, M., \& Hardy, E.\ 1991, 
			      \aj, 101, 94
\reference{dupuy1970}         du Puy, D.\ L.\ 1970, \aj, 75, 1143
\reference{fairall1977}       Fairall, A.\ P.\ 1977, \mnras, 180, 391
\reference{fason01991}        Fasano, G., \& Vio, R.\ 1991, \mnras, 249, 629
\reference{ferguson1989}      Ferguson, H.\ C., \& Sandage, A.\ 1989, ApJ, 
	                      346, L53
\reference{ferguson1994}      Ferguson, H.\ C., \& Binggeli, B.\ 
			      1994, \aap\ Rev., 6, 67
\reference{gondhaeker1984}    Gondhaeker, P.\ M., Morgas, D.\ H., 
			      Dopita, M., \& Phillip, A.\ P.\ 1984,
			      \mnras, 209, 59
\reference{gordon1981}        Gordon, D., \& Gottesman, S.\ T.\ 1981,
                              \aj, 86, 161
\reference{graham1982}        Graham, J.\ A.\ 1982, \pasp, 94, 244
\reference{haro1956}          Haro, G.\ 1956, Bol.\ Obs.\ Tomantzintla y
			      Tacubaya, 14, 8
\reference{hunter1985}        Hunter, D.\ A., \& Gallagher, J.\ S.\ 
			      1985, \apjs, 58, 533
\reference{ichikawa1989}       Ichikawa, S.-I.\ 1989, \aj, 97, 1600
\reference{ichikawa1986}      Ichikawa, S.-I., Wakamatsu, K.-I., \& 
	                      Okamura, S.\ 1986, \apjs, 60, 475
\reference{james1994}         James, P.\ A.\ 1994, \mnras, 269, 176
\reference{kunth1986}         Kunth, D., \& Sargent, W.\ L.\ W.\ 
			      1986, \aj, 91, 761
\reference{lambas1992}        Lambas, D.\ S., Maddox, S.\ J., \& 
			      Loveday, J.\ 1992, \mnras, 258, 404
\reference{landolt1992}       Landolt, A.\ U.\ 1992, \aj, 104, 340
\reference{loose1986}         Loose, H.\ H., \& Thuan, T.\ X.\ 1986,
			      in Star-Forming Dwarf Galaxies and
			      Related Objects, ed.\ D.\ Kunth, 
			      T.\ X.\ Thuan, \& T.\ T.\ Van
			      (Paris: Editions Fronti\`eres), 73
\reference{macalpine1981}     Macalpine, G.\ M., \& Williams, G.\ A.\ 1981,
			      \apjs, 45, 113
\reference{markarian1967}     Markarian, B.\ E.\ 1967, Astrofizika,
			      3, 55
\reference{maza1991}          Maza, J., Ruiz, M.\ T., Pena, M., 
		              Gonzalez, L.\ E.,\& Wischnjewsky, M.\ 
			      1991, \aaps, 89, 389
\reference{papaderos1996}     Papaderos, P., Loose, H.-M.,
			      Fricke, K.\ J., \& Thuan, T.\ X.\
			      1996,  \aap, 314, 59
\reference{ryden1992}         Ryden, B.\ S.\ 1992, \apj, 396, 445
\reference{ryden1996}         Ryden, B.\ S.\ 1996, \apj, 461, 146
\reference{ryden1994}         Ryden, B.\ S., \& Terndrup, D.\ M.\ 1994,
                              ApJ, 425, 43
\reference{ryden1997}         Ryden, B.\ S., Terndrup, D.\ M., Pogge, R.\ W.,
                              \& Lauer, T.\ R. 1997, \apj, submitted
\reference{sandage1984}       Sandage, A., \& Binggeli, B.\ 1984, \aj, 89, 919
\reference{searle1972}        Searle, L., \& Sargent, W.\ L.\ 1972,
                              \apj, 173, 25
%\reference{searle1973}        Searle, L., Sargent, W.\ L., \& Bagnuolo, 
%			      W.\ G.\ 1973, \apj, 179, 427
\reference{silk1987}          Silk, J., Wyse, R.\ F.\ G., \& Shields, 
			      G.\ A.\ 1987, \apj, 322, L59
\reference{staveley1992}      Staveley-Smith, L., Davies, R.\ D., \&
	                      Kinman, T.\ D.\ 1992, \mnras, 258, 334
\reference{terlevich1991}     Terlevich, R., Melnick, J.,
			      Masegosa, J., Moles, M.,
			      \& Copetti, M.\ V.\ F.\ 1991, \aaps, 91, 285 
\reference{thuan1983}         Thuan, T.\ X.\ 1983, \apj, 268, 667
\reference{thuan1991}         Thuan, T.\ 1991, in Proceedings of the 
			      Massive Stars in Starbursts, 
			      ed.\ C.\ Leitherer,  N.\ R.\ Walborn, 
			      T.\ M.\ Heckman, \& C.\ A.\ Norman 
			      (Cambridge: Cambridge University Press), 183
\reference{thuan1995}         Thuan, T.\ X., Izotov, Y. I., \&
                              Lipovetsky, V.\ A.\ 1995, ApJ, 445, 108
\reference{thuan1981}         Thuan, T.\ X., \& Martin, G.\ E.\ 1981, 
		              ApJ, 247, 823
\reference{tremblay}          Tremblay, B., \& Merritt, D.\ 1995, 
			      \aj, 110, 1039
\reference{wamsteker1985}     Wamsteker, W., Prieto, A., Vitores, A., 
			      Schuster, H.\ E., Danks, A.\ C., Gonzalez, R.,
			      \& Rodriguez, G.\ 1985, \aaps, 62, 255
\end{references}
\end{document}